\documentstyle[psfig,12pt,a4wide]{article}
\newcommand{\news}{\setcounter{equation}{0}}

\newcommand{\grad}{\mbox{$\nabla$}}
\def\eqn{\begin{equation}} 
\def\eeqn{\end{equation}}
\def\arr{\begin{array}} 
\def\earr{\end{array}}
\def\eqna{\begin{eqnarray}} 
\def\eeqna{\end{eqnarray}} 
\def\a{\alpha}
\def\b{\beta} 
\def\D{\Delta}
\def\s{\sigma} 
\def\d{\delta}

\def\r{\rho}

\def\th{\theta} 
\def\m{\mu} 
\def\n{\nu} 
\def\la{\lambda} 

\def\t{\tau} 
 
\def\g{\gamma}

\begin{document}

\vspace*{-.6in}
\thispagestyle{empty}
\begin{flushright}
PUPT-1947\\
DAMTP-2000-76\\
\end{flushright}

\vspace{.3in}
{\LARGE 
\begin{center}
{\bf Interacting Black Holes}
\end{center}}
\vspace{.3in}
\begin{center}
Miguel S. Costa\footnote{miguel@feynman.princeton.edu}\\
\vspace{.1in}
\emph{Joseph Henry Laboratories\\ Princeton University \\ 
Princeton, New Jersey 08544, USA}\\
\vspace{.3in}
Malcolm J. Perry\footnote{malcolm@damtp.cam.ac.uk}\\
\vspace{.1in}
\emph{D.A.M.T.P., Center for Mathematical Sciences\\ 
University of Cambridge\\
Wilberforce Road, Cambridge CB3 0WA, UK}
\end{center}

\vspace{.5in}

\begin{abstract}
We revisit the geometry representing $l$ collinear Schwarzschild black holes.
It is seen that the black holes' horizons are deformed by their mutual
gravitational attraction. The geometry has a string like conical
singularity that connects the holes but has nevertheless a well defined action. 
Using standard gravitational thermodynamics techniques we determine
the Free energy for two black holes at fixed temperature and distance, their
entropy and mutual force.
When the black holes are far apart the results agree with 
Newtonian gravity expectations. 
This analyses is generalized to the case of charged black holes. 
Then we consider black holes embedded in String/M-theory as bound states of branes. 
Using the effective string description of these bound states and for large separation 
we reproduce exactly the semi-classical result for the entropy, including
the correction associated with the interaction between the holes.
\end{abstract}
\newpage

\section{Introduction}
\news

A longstanding and poorly explored problem in General Relativity is the 
interaction between Schwarzschild black holes, or more generally the interaction 
between Reissner-Nordstrom black holes. While extremal Reissner-Nordstrom black 
holes are in equilibrium, corresponding in the modern language to BPS states
of an yet unknown supersymmetric theory, in the non-extremal case the holes
are expected to attract each other. It may then come as a surprise that a static
solution representing $l$ Schwarzschild black holes placed along a common axis is 
indeed known in the literature \cite{IsraKhan}. 
The aim of this paper is to study such holes 
by using standard gravity techniques as well as the modern String Theory 
approach to black hole physics. 

The geometry representing $l$ collinear Schwarzschild black holes has been known
for a while \cite{IsraKhan}. It belongs to the general class of static
axisymmetric Einstein vacua studied by Weyl \cite{Weyl}, who showed that the 
problem of finding such solutions can be reduced to an associated problem 
in Newtonian gravity. In the case of $l$ collinear Schwarzschild black holes of
mass $M_n$ each, one needs to find the classical potential for a system of $l$ rods 
each with mass $M_n$ and length $2M_nG$. It may appear that for more than one hole this
solution represents $l$ Schwarzschild black holes in equilibrium. However, such a
solution fails the elementary flatness requirement \cite{EinsRosen}. In other words,
the geometry contains conical singularities (similar to those found for cosmic strings) 
on the sections between the holes. The failure to find a spacetime free of 
singularities is interpreted as an instability, i.e. two of such black holes 
cannot be in static equilibrium \cite{HagenSeif,Gibb}. This fact may seem discouraging,
however the above solution, and in particular the contribution from the conical 
singularities, still has a well defined gravitational action \cite{GibbPerry}.
This means that we can use standard Euclidean Gravity thermodynamical arguments.
The corresponding thermodynamical potential describes a system of black holes at fixed 
temperature and at fixed distance, i.e.  we interpret the conical singularities as 
boundary conditions that keep the holes at fixed distance preventing the collapse of
the system. 

A similar analyses can be done for charged black holes. The motivation
is that in some cases such holes have a natural embedding in 
String/M-theory as bound states of branes. Here we shall consider the
triple intersection of M5-branes over 
a string \cite{KlebTsey,CvetTsey}. Remarkably, using the
effective string description of these bound states and for large separation between the
holes we were able to reproduce the entropy formula derived in the gravitational analysis. 
We find this fact 
reassuring, confirming that the old solution representing collinear Schwarzschild 
black holes and the new charged ones are in fact sensible solutions, if interpreted
appropriately.

This paper starts in Section Two by reviewing the geometry representing $l$ collinear 
Schwarzschild black holes. Then we analyze the black holes' horizons, in particular how 
they are deformed by the mutual gravitational attraction between the holes. 
We shall see that the horizons 
area and surface gravity are changed by the presence of other holes. Then we
discuss the conical singularity between the holes and its stringy description. The 
energy associated with such string is identified with the binding energy between 
the black holes and agrees with Newtonian gravity for large distances. In fact, we shall
argue that for short distances, the gravitational approach breaks down because we
are dealing with physics at the Planck scale. This happens when the holes are
expected to merge.
We proceed by determining the action for the Euclidean solution which
is related to the Free energy for a system of black holes at fixed temperature
and at fixed distance. We find that for large separation the entropy follows the 
area law and the generalized force 
between the holes is in agreement with Newtonian gravity.

Section Three is devoted to the study of charged black holes. First we embed the 
collinear Schwarzschild solution in five dimensions, apply a boost along the 
extra compact direction and reduce back to four dimensions. This procedure 
generates a solution representing $l$ collinear charged black hole in 
Einstein-Maxwell gravity with a 
scalar field. Then it is straightforward to find the solution with 
arbitrary $e^{a\phi}F^2$ coupling between the scalar field and the 
electromagnetic field. We analyze the properties of the charged black holes, 
with an emphasis on thermodynamics, 
generalizing the results mentioned above for uncharged black holes.

After finding the solution representing $l$ collinear non-extremal charged
black holes we were
lead to use String Theory to reproduce the results obtained in the semi-classical
approach. This is the topic discussed in Section Four.
We shall consider two black holes, each embedded in String/M-theory as 
a triple intersection of M5-branes over a string. Using the effective string description
of such black holes we were able to reproduce exactly the corrections to the entropy
due to the interaction between the holes. Furthermore, we resolve a puzzle that arose
in the semi-classical calculation of the mean energy. 
Apparently the term associated with the interaction 
between the holes has the wrong sign to be associated with a Coulomb attractive
potential. Careful considerations of the effective string model lead to the following
solution of the problem: the left- and right-movers on the effective strings interact 
through an attractive Coulomb potential reducing their energy by $V_{int}$.
As a consequence the left- and right-mover levels are shifted, 
leading to the correct entropy
and energy formulae. Essentially the problem arises because Newtonian gravity ignores
the black holes internal degrees of freedom, which are affected by this long
range interaction.

We give our conclusions in Section Five.

\section{Collinear Schwarzschild Black Holes}
\news

Any static axisymmetric vacuum space-time, with two commuting killing vectors
$\frac{\partial\ }{\partial t}$ and $\frac{\partial\ }{\partial\phi}$, 
possesses a metric that can be written in the form
\eqn
ds^2=-Vdt^2+V^{-1}\g_{ij}dx^idx^j\ ,
\label{metric}
\eeqn
with
\eqn
\g_{ij}dx^idx^j=e^{2K}\left( d\r^2+dz^2\right) +\r^2d\phi^2\ .
\label{spatial}
\eeqn
The functions $V$ and $K$ depend on $\r$ and $z$ only. With these assumptions
the Einstein field equations read
\eqn
K_{\r}=\frac{\r}{4V^2}\left(V_{\r}^{\ 2}-V_{z}^{\ 2}\right)\ ,\ \ \ \ \ \ 
K_{z}=\frac{\r}{2V^2}V_{\r}V_z\ ,
\label{eqsofmotion}
\eeqn
with the integrability condition $V\grad^2V=\grad V\cdot\grad V$, where $\grad$ is the 
taken with respect to flat Euclidean space in cylindrical coordinates. The subscripts 
in the functions $V$ and $K$ represent partial differentiation with respect to that
variable. Defining a function $U$ by $V=e^{2U}$ the equations of motion become
\eqn
K_{\r}=\r\left(U_{\r}^{\ 2}-U_{z}^{\ 2}\right)\ ,\ \ \ \ \ \ 
K_{z}=2\r\,U_{\r}U_z\ ,
\label{eqsofmotion2}
\eeqn
with the integrability condition
\eqn
\grad^2U=\left(\frac{\partial^2\ }{\partial\r}+
\frac{1}{\r}\frac{\partial\ }{\partial\r}
+\frac{\partial^2\ }{\partial z^2}\right)U=0\ .
\label{Newton}
\eeqn
The function $U$ is formally the Newtonian potential outside some mass distribution. 
Thus the problem of finding the metric for a static axisymmetric space-time has been
reduced to an associated problem in Newtonian gravity. 

In the case of the Schwarzschild black hole $U$ is the potential for a rod of 
length $\m=2MG$ and mass $M$ (i.e. mass density $1/2G$). The solution
representing $l$ collinear Schwarzschild black holes is obtained by considering
the potential for $l$ rods of length $\m_n=2M_nG$ and mass $M_n$ \cite{IsraKhan}:
\eqn
\arr{c}
\displaystyle{U=\frac{1}{2}\sum_{n=1}^{l}
\ln\frac{r_n^++r_n^--\m_n}{r_n^++r_n^-+\m_n}}\ ,\\\\
\displaystyle{K=\frac{1}{4}\sum_{n,m=1}^{l}
\ln \left(\frac{r_n^+r_m^-+[z-(z_n+\m_n/2)][z-(z_m-\m_m/2)]+\r^2}
{r_n^+r_m^++[z-(z_n+\m_n/2)][z-(z_m+\m_m/2)]+\r^2}\right.}\\\\
\displaystyle{\left.\ \ \ \ \ \ \ \ \ \ \ \ \ \ \ \ \ \ \ \ \ \ \ 
\frac{r_n^-r_m^++[z-(z_n-\m_n/2)][z-(z_m+\m_m/2)]+\r^2}
{r_n^-r_m^-+[z-(z_n-\m_n/2)][z-(z_m-\m_m/2)]+\r^2}
\right)}\ ,
\earr
\label{solution}
\eeqn
where $z_n$ is the $n$-th rod mid point and 
$r_n^+$ and $r_n^-$ are given by
\eqn
r_n^{\pm}=\sqrt{\r^2+[z-(z_n\pm \m_n/2)]}\ .
\eeqn
For simplicity we shall consider the case with two black holes but 
the results here presented are easily generalized to an arbitrary number of holes.
In Figure 1 we represent the rods in the coordinate system used so far.  
Notice that in the single rod case the change of coordinates
\eqn
\r=r\sqrt{1-\frac{\m_1}{r}}\sin{\th}\ ,\ \ \ \ \ \ 
z-z_1=(r-\m_1/2)\cos{\th}\ .
\label{Schwvar}
\eeqn
brings the metric to the usual form known for the Schwarzschild black hole.

\begin{figure}
\centerline{\psfig{figure=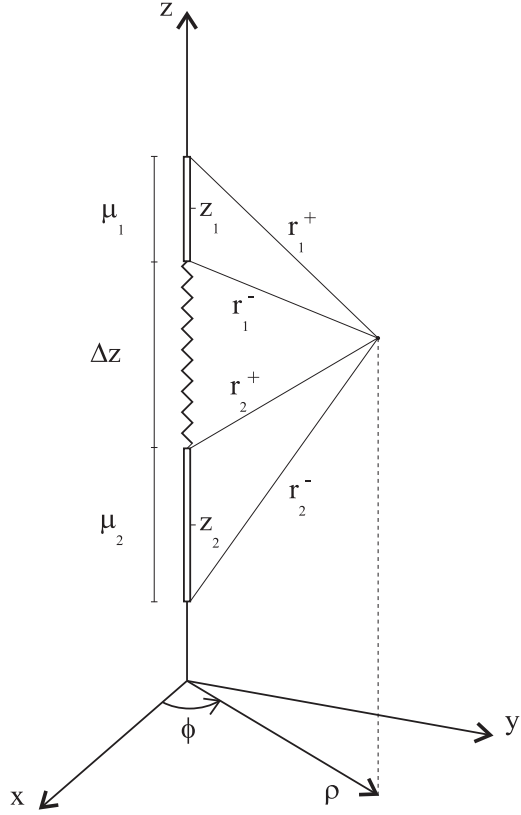,width=2.7in}}
\caption{\small{The case with two Schwarzschild black holes. Each rod is
the locus for the holes' horizon and the section between the rods has a conical 
singularity. The Newtonian distance between the rods is $\D z=z_1-z_2-(\m_1+\m_2)/2$.}}
\label{fig1}
\end{figure}

Let us start by determining the ADM mass associated with this background. From the
asymptotic behavior 
\eqn
V\sim 1-\frac{\m_1+\m_2}{r}+{\cal O}\left(\frac{1}{r^2}\right)\ ,\ \ \ \ 
K\sim \frac{1}{4}\ln \left(1+{\cal O}\left(\frac{1}{r^2}\right)\right)\sim
{\cal O}\left(\frac{1}{r^2}\right)\ ,
\label{asympt}
\eeqn
we conclude that the mass is $M_1+M_2$ as expected. Asymptotically space-time is flat
and, provided the polar angle $\phi$ is defined modulo $2\pi$, free of any singularity. 

Next consider the metric near the $z$ axis. Along the sections that extend from the
rods to infinity we have $K=0$ and $V$ has a well behaved limit. 
Hence the metric is well behaved
along these sections of the $z$ axis. The holes' horizons correspond to the sections
with rods. We postpone a detail discussion of the horizons to the next subsection. 
Along the section between both rods the function $K$ is
\eqn
K=\ln\left(\frac{\D z+\m_1+\m_2}{\D z+\m_1}\frac{\D z}{\D z+\m_2}\right)\ ,
\label{K}
\eeqn
and $V$ has a well behaved limit as well. Inspecting the spatial part of the
metric given by equation (\ref{spatial}) we conclude that there is a conical 
singularity connecting both rods. The deficit angle $\d$ satisfies
\eqn
\frac{\d}{2\pi}=\left(1-e^{-K}\right)=
-\frac{\m_1\m_2}{(\D z +\m_1+\m_2)\D z}\ .
\label{defangle}
\eeqn
In fact $\d$ is negative so we have an excess angle. We conclude that the solution
representing two Schwarzschild black holes is not free of singularities. 
This conical singularity signals an instability: two Schwarzschild black holes
cannot be in equilibrium \cite{HagenSeif,Gibb}. 
This fact has discouraged further work on the subject
over the years. However, it was pointed out in \cite{GibbPerry} that this solution
has a well defined gravitational action. Essentially this means that we can do 
gravitational thermodynamics and, if correctly interpreting the conical singularity,
extract new results from an apparently ill defined geometry. We end here our review
of collinear Schwarzschild black holes. Further details can be found in references
\cite{IsraKhan,HagenSeif,Gibb}.

\subsection{Horizons}

To study the horizons we need to investigate the behavior of the metric near the rods. 
To be definite we shall consider the first rod corresponding to the black hole with
mass $M_1$. Then performing the coordinate transformation (\ref{Schwvar}) the
metric near $r=\m_1=2M_1G$ takes the form
\eqn
\arr{l}
\displaystyle{ds^2\sim
f^2(\th)
\left[-\left(1-\frac{\m_1}{r}\right)dt^2
+\left(\frac{\D z+\m_1+\m_2}{\D z+\m_1}\right)^2
\left(1-\frac{\m_1}{r}\right)^{-1}dr^2
\right]}\\\\
\displaystyle{\ \ \ \ \ \ \ \ \ \ \ +
\m_1^{\ 2}\left[\left(\frac{\D z+\m_1+\m_2}{\D z+\m_1}\right)^2
f^2(\th)\,d\th^2+
f^{-2}(\th)\sin^2{\th}\,d\phi^2\right]}\ ,
\earr
\label{horizon}
\eeqn
where
\eqn
f(\th)=\sqrt{\frac{\D z+\m_1(\cos{\th}+1)/2}{\D z+\m_1(\cos{\th}+1)/2+\m_2}}\ .
\eeqn
Clearly the surface $r=\m_1$ and $t=const.$ is a horizon with surface gravity and
area given by
\eqn
\arr{c}
\displaystyle{\kappa_1=
\frac{1}{2\m_1}\,\frac{\D z+\m_1}{\D z +\m_1+\m_2}}\ ,\\\\
\displaystyle{A(H_1)=
4\pi\m_1^{\ 2}\,\frac{\D z +\m_1+\m_2}{\D z+\m_1}}\ .
\earr
\label{kA}
\eeqn
As usual the temperature of the black hole is related to the horizon surface gravity
by $T_1=\kappa_1/(2\pi)$. In the limit $\D z=0$, we have $\kappa^{-1}=2(\m_1+\m_2)$
and $A(H_1+H_2)=4\pi(\m_1+\m_2)^2$. This limit corresponds to the case when
both rods come together forming a single rod with mass $M_1+M_2$ and therefore 
a single black hole. On the other hand in the limit $\D z\rightarrow\infty$, we 
have $\kappa_n^{\ -1}=2\m_n$ and $A(H_n)=4\pi\m_n^{\ 2}$. This is expected because 
when both holes are far apart each one should behave as in the single black 
hole case. 

We shall be interested in the study of
interacting black holes in thermal equilibrium.
Equating the holes' temperature and discarding the trivial $\D z=0$ case we must 
have $M_1=M_2\equiv M$ (or $\m_1=\m_2\equiv\m$). 
We are led to the following picture. For large $\D z$
we have two holes with mass $M$ and temperature $T=(8\pi MG)^{-1}$. The horizons
are almost spherical and the holes are connected by a conical singularity. As 
$\D z$ decreases we expect the horizons to be deformed to a egg like shape, with
the conical singularity that connects both holes emanating from the flat sides of the
horizons. The temperature decreases. When $\D z=0$ the horizons become the North and 
South hemispheres of a single black hole horizon with a mass equal to $2M$. 

To make the above claim precise consider the angular part of the 
metric near the first rod
\eqn
ds_{ang}^2\sim
\m^2\left[ A^2f^2(\th)\,d\th^2+f^{-2}(\th)\sin^2{\th}\,d\phi^2\right]\ ,
\label{anghorizon}
\eeqn
with
\eqn
A=\frac{\D z+2\m}{\D z+\m}\ , \ \ \ \ \ 
f(\th)=\sqrt{\frac{\D z+\m(\cos{\th}+1)/2}{\D z+\m(\cos{\th}+3)/2}}\ .
\eeqn
Next change to the angular variable $\tilde{\th}=\tilde{\th}(\th)$ by
requiring
\eqn
ds_{ang}^2\sim
\m^2g^2(\tilde{\th})\left[d\tilde{\th}^2+\sin^2{\tilde{\th}}\,d\phi^2\right]\ .
\label{anghorizon2}
\eeqn
Then $\tilde{\th}=\tilde{\th}(\th)$ and $g(\tilde{\th})$ are uniquely determined by
\eqn
\int\frac{d\tilde{\th}}{\sin{\tilde{\th}}}=A\int f^2(\th)\frac{d\th}{\sin{\th}}\ ,
\ \ \ \ \ g(\tilde{\th})\sin{\tilde{\th}}=\frac{\sin{\th}}{f(\th)}\ .
\eeqn
The integral can be done explicitly with the result
\eqn
\tan{(\tilde{\th}/2)}=C\tan{(\th/2)}
\left(\cos{(\th/2)}\right)^{\frac{4}{(\a +2)^2}}
\left(3+\cos{\th}+\a\right)^{-\frac{2}{(\a +2)^2}}\ ,
\eeqn
where $C$ an integration constant and $\a=2\D z/\m$. The metric (\ref{anghorizon2})
defines an axisymmetric two-surface with radius $R(\tilde{\th})=\m\,g(\tilde{\th})$.
The constant $C$ can be fixed by requiring $g'(0)=0$, i.e. the surface is
regular at the pole opposite to the  other hole. The result is
\eqn
C^2=\frac{\a^2+6\a+4}{\a+2}(\a+4)^{\frac{4}{(\a+2)^2}-1}\ .
\eeqn
As a check notice that for $\D z=0$, $C=\sqrt{2}$ and 
$\tan{(\tilde{\th}/2)}=\sqrt{2}\sin{(\th/2)}/\sqrt{3+\cos{\th}}$.
In this case $\tilde{\th}\in [0,\pi/2]$ and the two-surface is a North hemisphere 
with radius $2\m$ as pointed out before. 
When $\D z\rightarrow\infty$ we have $C=1$, $\tilde{\th}=\th$ and the two-surface 
becomes a two-sphere with radius $\m$. In Figure 2 the horizons are represented
for the cases $\a=10,1,0.1$.

\begin{figure}
\centerline{\psfig{figure=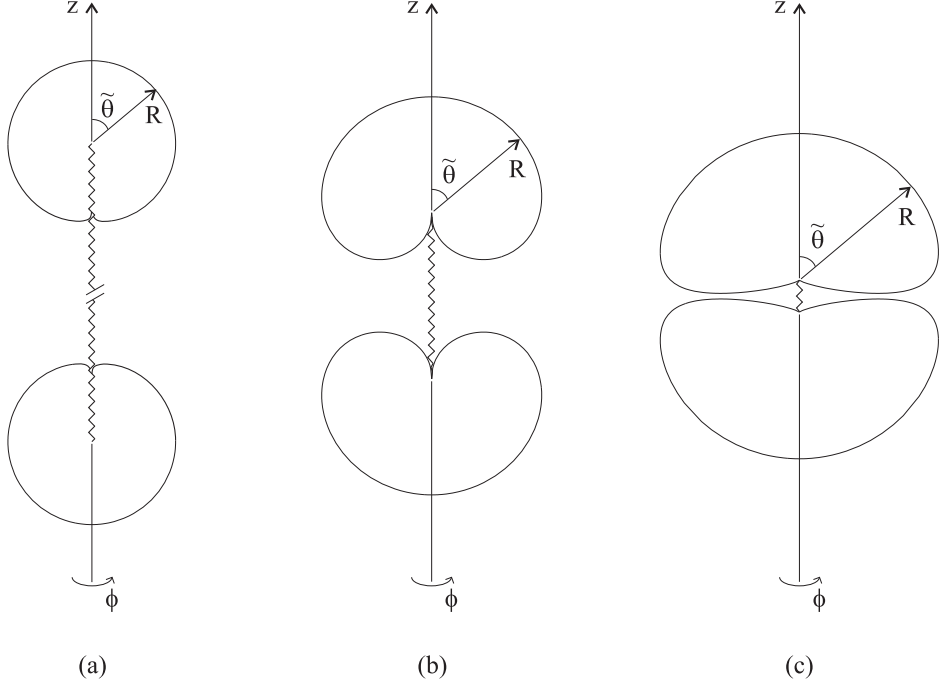,width=4.9in}}
\caption{\small{The horizons of the black holes $R=R(\tilde{\th})$ for different values of
the distance parameter $\a=2\D z/\m$. For (a) large separation $\a=10$ the horizons 
are almost spherical with radius $R\sim 2MG$. The proper distance between the holes 
is $L\sim (1.83)\D z$. For (b) $\a=1$ the horizons are quite deformed and 
$L\sim (4.32)\D z$. For (c) small separation $\a=0.1$ each horizon becomes a 
hemisphere with radius $R\sim 4MG$ and
$L\sim (5.99)\D z$.}}
\label{fig2}
\end{figure}

Finally we investigate the relation between the proper length $L$ of the section with a
conical singularity, i.e. the distance  between the singular points in the holes' 
horizons, and the distance $\D z$ in the associated Newtonian problem. 
A short calculation shows that
\eqn
L=e^K\int_{z_2+\m/2}^{z_1-\m/2}\frac{dz}{\sqrt{V}}=
\D z\left(\frac{\a +4}{\a +2}\right)^2E(m)\ ,\ \ \ \ 
m=\left(\frac{\a}{\a +4}\right)^2\ ,
\eeqn
where $E(m)$ is the Complete Elliptic Integral of the Second Kind. Indeed only for 
large distances between the holes we have $L\sim \D z$ (recall that $E(1)=1$). 
For small distances we have $L\sim 2\pi\D z$ (recall that $E(0)=\pi/2$).

\subsection{Conical Singularity}

Let us start by writing the spacetime metric along the section between both rods
\eqn
ds^2\sim \left( -V(z)\,dt^2+V^{-1}(z)\,e^{2K}dz^2\right)
+V^{-1}(z)\left( e^{2K}d\r^2+\r^2d\phi^2\right)\ ,
\label{rod}
\eeqn
where
\eqn
V(z)=\frac{z-(z_1-\m/2)}{z-(z_1+\m/2)}\,\frac{z-(z_2+\m/2)}{z-(z_2-\m/2)}\ ,\ \ \ \ 
e^K=\frac{(\D z+2\m)\D z}{(\D z+\m)^2}\ .
\eeqn
The deficit angle along the axis is
\eqn
\frac{\d}{2\pi}=-\frac{\m^2}{(\D z+2\m)\D z}\ .
\eeqn
This geometry is Ricci flat, however the presence of a conical singularity on the
axis means that \cite{Regge} 
\eqn
\frac{1}{2}\int d^4x\sqrt{-g}R= Area\cdot \d\ ,
\eeqn
where $Area$ is the space-time area of the surface spanned by the conical singularity. 
It follows that the Ricci scalar can be defined distributionally according to
\eqn
R=-V(z)\frac{4\pi\m^2}{(\D z+2\m)\D z}\,\d(x,y)\ ,
\eeqn
where $(x,y)$ are the cartesian coordinates in the plane orthogonal to the $z$ axis. 
Furthermore, since the four-dimensional metric factorizes and this plane is a 
two-dimensional manifold we can use the result 
$R_{ab}=(R/2)g_{ab}$ to determine the non-zero components of the Ricci tensor:
$R_{xx}=R_{yy}=R/2$.
This defines, by means of the Einstein field equations $G_{\m\n}=8\pi G\,T_{\m\n}$,
an effective energy-momentum tensor associated with the conical singularity:
\eqn
T^{\m\n}=-\frac{\m^2}{4G(\D z+2\m)\D z}\,\d(x,y)\,
{\rm diag}\left(1,0,0,-V^2e^{-2K}\right)\ .
\label{emt}
\eeqn
The situation is similar to the case of cosmic strings, but with the 
important difference that the deficit angle and therefore the tension 
$T=-\m^2/(4G(\D z+2\m)\D z)$ are negative. The singularity is therefore usually 
referred to as a strut rather than a string.

To determine the energy associated with this strut, let 
$v=\frac{\partial\ }{\partial t}$ and define the unit vector $\zeta$ by
$\zeta=v/(-v\cdot v)^{1/2}$. 
Then the energy density as seen by a static observer
placed at infinity is
\eqn
T^{\m\n}v_{\m}\zeta_{\n}=-\frac{\m^2V^{3/2}(z)}{4G(\D z+2\m)\D z}\,\d(x,y)\ .
\eeqn
The total energy associated with the strut as seen by this observer is obtained 
by integrating over the spatial directions 
\eqn
E=-\frac{\m^2\D z}{4G(\D z+\m)^2}\ .
\eeqn
This is interpreted as the gravitational binding energy between both black
holes. In fact, for $\m/\D z\ll 1$ we have $E\sim-M^2G/\D z$, the Newtonian
potential between two particles of mass $M$. 

Next let $w=\frac{\partial\ }{\partial z}$ and define the unit vector $\la$ by
$\la=w/(w\cdot w)^{1/2}$. The pressure along the $z$ axis is
\eqn
T^{\m\n}\la_{\m}\la_{\n}=\frac{\m^2V(z)}{4G(\D z+2\m)\D z}\,\d(x,y)\ ,
\eeqn
which equals minus the local energy density. The pressure on the strut is
obtained by integrating over the $x-y$ plane:
\eqn
P=\frac{\m^2}{4G(\D z+2\m)\D z}\ ,
\eeqn
which for $\m/\D z\ll 1$ gives $P\sim M^2G/\D z^2$. This explains why the black 
holes are kept at a fixed distance since the pressure exerted on the holes will
cancel the gravitational attraction. 

To understand the physics underlying this problem, notice that this geometry was
found by requiring the coordinate distance  $\D z$ to be fixed. In fact, in the
associated Newtonian problem we considered the potential for two rods 
separated by a distance $\D z$ but the gravitational attraction between the
rods was ignored. Certainly the black holes will attract each other and
are not in equilibrium. This is the reason why requiring $\D z$ fixed leads to 
a conical singularity connecting both holes. Effectively this conical singularity 
can be seen as a strut connecting both holes. The strut energy is interpreted 
as the interaction energy between the black holes while its pressure prevents 
the gravitational collapse of the system. For large distances between the holes
this interpretation is consistent with Newtonian gravity. Hence the
presence of the conical singularity is in agreement with the energetics of the 
problem  and with keeping $\D z$ fixed. 

\subsection{Gravitational Thermodynamics}

The novelty of this solution is that its action is well defined.
Therefore by the usual semi-classical gravitational thermodynamics arguments we
can calculate the Free energy for two Schwarzschild black holes at fixed temperature
and distance. This defines the entropy of the system and the 
generalized force conjugate to $\D z$, which for large distances should agree
with the Newton force between two particles of mass $M$. Furthermore, the system
will evolve such that the minimum of the thermodynamical potential is attained.

Before we start this calculation it is important to analyze whether the 
semi-classical evaluation of the partition function is reliable. 
We expect the General Relativity 
description to be valid for small deficit angle $|\d|$. Effectively this
corresponds to have a strut tension well below the Planck scale:
\eqn
\frac{\m^2}{4G(\D z+2\m)\D z}\ll \frac{1}{G}\Rightarrow 
\frac{\m}{\D z}\ll 1\ ,
\eeqn
which happens when the black holes are at a large proper distance 
from each other. Thus this work will focus on well separated black holes 
interacting through a Coulomb potential. This corresponds to the limit where both
holes' horizons are nearly spherical as represented in Figure 2(a).

The Euclidean action for the solution is
\eqn
I=\b\left( 2\,\frac{M}{2}+\frac{M^2G\D z}{(\D z+2MG)^2}\right)
=\b\left( \frac{\m}{2G}+\frac{\m^2\D z}{4G(\D z+\m)^2}\right)\ ,
\label{action}
\eeqn
where 
\eqn
\frac{1}{T}=\b=4\pi\m\,\frac{\D z+2\m}{\D z+\m}\ ,
\label{invtemp}
\eeqn
is the Euclidean time periodicity. The metric is regular at the fixed point
sets of the Euclidean time translation Killing vector 
$\frac{\partial\ }{\partial\t}$, that correspond to the two horizons in the 
Lorentzian version. These bolts are connected by the conical singularity along
which $\frac{\partial\ }{\partial\t}$ has no fixed points. 

From (\ref{action}) we can read the thermodynamical potential for two 
Schwarzschild black holes at fixed temperature and distance evaluated in 
the semi-classical limit:
\eqn
{\cal F}(T,\D z)=\frac{\m}{2G}+\frac{\m^2\D z}{4G(\D z+\m)^2}\ ,
\label{free}
\eeqn
where $\m=\m(T,\D z)$ through the relation (\ref{invtemp}). Then the entropy and
generalized force conjugate to $\D z$ are 
\eqn
S=-\left.\frac{\partial {\cal F}}{\partial T}\right|_{\D z}\ ,\ \ \ \ 
F_{\D z}=-\left.\frac{\partial {\cal F}}{\partial \D z}\right|_T\ .
\eeqn
It is useful to determine first 
$\left.\frac{\partial \m}{\partial T}\right|_{\D z}$ and
$\left.\frac{\partial \m}{\partial \D z}\right|_T$. The result is
\eqn
\arr{c}
\displaystyle{
\left.\frac{\partial \m}{\partial T}\right|_{\D z}=
-4\pi\m^2\frac{(\D z +2\m)^2}{\D z^2+4\m\D z+2\m^2}=
-4\pi\m^2\left[ 1+{\cal O}\left(\frac{\m}{\D z}\right)^2\right]}\ ,\\\\
\displaystyle{
\left.\frac{\partial \m}{\partial \D z}\right|_T=
\frac{\m^2}{\D z^2+4\m\D z+2\m^2}=
\frac{\m^2}{\D z^2}\left[ 1+{\cal O}\left(\frac{\m}{\D z}\right)\right]}\ .
\earr
\eeqn
Then the entropy and generalized force are 
\eqn
\arr{c}
\displaystyle{
S=8\pi GM^2\left[ 1+\frac{\m}{\D z}+
{\cal O}\left(\frac{\m}{\D z}\right)^2\right]=
\frac{1}{4G}\left( A(H_1^{\phantom{1}})+A(H_2^{\phantom{1}})\right)}\ ,\\\\
\displaystyle{
F_{\D z}=-\frac{M^2G}{\D z^2}
\left[ 1+{\cal O}\left(\frac{\m}{\D z}\right)\right]}\ .
\earr
\eeqn
To order $\m/\D z$ the entropy satisfies the area law. The correction arises from
the conical singularity contribution to the Free energy and it is associated to
the deformation of the horizon due to the gravitational attraction between the
holes. The generalized force $F_{\D z}$ matches exactly the Newtonian force
between two particles of mass $M$, as expected for largely separated black holes.

In spite of the fact that the above results seem consistent with Newtonian gravity
it may look surprising that for large distances the second term in the Free energy
(\ref{free}) has the wrong sign to be interpreted as the Free energy for two gravitating
particles of mass $M$. The same comment applies to the mean value of the energy
\eqn
E={\cal F}+TS=2M+\frac{M^2G}{\D z}
\left[ 1+{\cal O}\left(\frac{\m}{\D z}\right)\right]\ .
\label{energy}
\eeqn
Firstly, notice that the Free energy decreases as $\D z$ decreases because $M$
is not fixed. Figure 3 shows a plot of the Free energy for fixed temperature as 
$\D z$ varies.
The minimum of the thermodynamical potential corresponds to the single black hole
case while for large distances the holes will tend to attract each other. 
\begin{figure}
\centerline{\psfig{figure=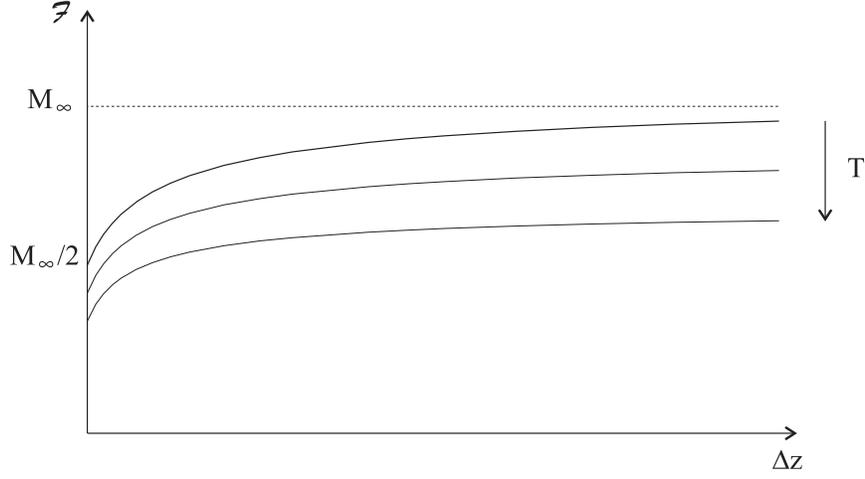,width=4.5in}}
\caption{\small{The Free Energy as a function of the distance parameter $\D z$ for
fixed temperature. The mass of the holes when they are far apart
is related to the temperature by $M_{\infty}=(8\pi GT)^{-1}$. For $\D z=\infty$
we have two holes of mass $M_{\infty}$ and a Free Energy 
${\cal F}=2(M_{\infty}/2)$, while for $\D z=0$ we have one hole with the same mass 
and a Free Energy ${\cal F}=M_{\infty}/2$. To minimize the thermodynamical
potential the holes tend to decrease their distance (and to increase the temperature 
as usual for Schwarzschild black holes).}}
\label{fig3}
\end{figure}
Secondly, the apparently wrong signs in (\ref{free}) and (\ref{energy}) may be
telling us that there are other internal degrees of freedom in the black holes
that are being excited due to the gravitational interaction, with this net result. 
With the help of String Theory we shall see in Section Four that this is indeed 
the case. In fact, only by fixing the internal degrees of freedom of the black
holes we should expect the variation of the energy to follow the Newtonian gravity
prediction. To see this consider the energy variation under an adiabatic 
transformation 
\eqn
dE=TdS-F_{\D z}\,d(\D z)\ .
\eeqn
For a reversible process, i.e. at fixed entropy keeping the number of internal 
degrees of freedom fixed, we have
\eqn
dE=\frac{M^2G}{\D z^2}\,d(\D z)\Rightarrow
E_f-E_i=-\left.\frac{M^2G}{\D z}\right|^{\D z_f}_{\D z_i}\ ,
\eeqn
in agreement with Newtonian gravity.

\section{Charged Black Holes}
\news

The goal of this section is to generalize the previous results to charged black
holes. First consider the five-dimensional Einstein theory and compactify the 
fifth direction on a circle according to
\eqn
ds_5^{\ 2}=e^{-\frac{2}{\sqrt{3}}\phi}\left(dy-A_{\m}dx^{\m}\right)^2
+e^{\frac{1}{\sqrt{3}}\phi}g_{\m\n}dx^{\m}dx^{\n}\ .
\label{5Dmetric}
\eeqn
Then the action for the four-dimensional theory reads
\eqn
I=\frac{1}{16\pi G}\int d^4x\sqrt{-g}
\left[R-\frac{1}{2}(\partial\phi)^2-\frac{1}{4}e^{-\sqrt{3}\phi}F^2\right]\ ,
\eeqn
where $F=dA$. Next embed the solution representing $l$ collinear Schwarzschild
black holes (\ref{metric}) in five dimensions by adding an extra flat direction
\eqn
ds_5^{\ 2}=-Vdt^2+V^{-1}\g_{ij}dx^idx^j+dy^2\ ,
\eeqn
where $\g_{ij}$ is given by (\ref{spatial}) and the functions $V=e^{2U}$ and $K$ are as
in (\ref{solution}). Applying a boost along the $y$ direction the metric becomes
\eqn
\arr{l}
\displaystyle{ds_5^{\ 2}=
-\frac{V}{\cosh^2{\s}-V\sinh^2{\s}}\,dt^2+V^{-1}\g_{ij}dx^idx^j}\\
\displaystyle{\ \ \ \ \ \ \ \ 
+\left(\cosh^2{\s}-V\sinh^2{\s}\right)
\left(dy-\frac{(1-V)\cosh{\s}\sinh{\s}}{\cosh^2{\s}-V\sinh^2{\s}}dt\right)^2}\ .
\earr
\eeqn
According to the ansatz (\ref{5Dmetric}) this corresponds to the following 
solution of the four-dimensional theory
\eqn
\arr{c}
\displaystyle{
ds^2=-H^{-\frac{1}{2}}Vdt^2+H^{\frac{1}{2}}V^{-1}\g_{ij}dx^idx^j}\ ,\\
\displaystyle{
F=-\coth{\s}\,d(H^{-1})\wedge dt\ ,\ \ \ 
e^{-\frac{2}{\sqrt{3}}\phi}=H}\ ,
\earr
\eeqn
where $H=1+\sinh^2{\s}\,(1-V)$.

The above Ka\l u\.{z}a-Klein solution can be generalized to an arbitrary 
$e^{a\phi}F^2$ coupling of the scalar field to the gauge field. The action
\eqn
I=\frac{1}{16\pi G}\int d^4x\sqrt{-g}
\left[R-\frac{1}{2}(\partial\phi)^2-\frac{1}{4}e^{a\phi}F^2\right]\ ,
\eeqn
has the following solution representing $l$ collinear electrically charged black
hole
\eqn
\arr{c}
\displaystyle{
ds^2=H^{\frac{2}{1+a^2}}\left[-\frac{V}{H^{\frac{4}{1+a^2}}}dt^2+
V^{-1}\g_{ij}dx^idx^j\right]}\ ,\\
\displaystyle{
F=-\frac{2}{\sqrt{1+a^2}}\coth{\s}\,d(H^{-1})\wedge dt\ ,\ \ \ 
e^{2\phi}=H^{\frac{4a}{1+a^2}}}\ .
\earr
\eeqn
The dual magnetic solution can be found by transforming $\phi\rightarrow -\phi$ and
$F\rightarrow \star F$. For the sake of clarity we shall consider only electrically 
charged black holes.

From the asymptotics of this solution we can determine the ADM mass, the electric
charge and the charge associated with the scalar field:
\eqn
\arr{c}
\displaystyle{M_{total}=\sum_{n=1}^{l}M_n=
\frac{1}{2G}\left( 1+\frac{2}{1+a^2}\sinh^2{\s}\right)\sum_{n=1}^{l}\m_n}\ ,\\
\displaystyle{Q_{total}=\sum_{n=1}^{l}Q_n=
\frac{1}{\sqrt{1+a^2}}\sinh{(2\s)}\sum_{n=1}^{l}\m_n}\ ,\\
\displaystyle{\Sigma_{total}=\sum_{n=1}^{l}\Sigma_n=
\frac{2a}{1+a^2}\sinh^2{\s}\sum_{n=1}^{l}\m_n}\ .
\earr
\eeqn
The mass and charges of the $n$-th black hole together with the non-extremality 
parameter $\m_n$ satisfy the important relation
\eqn
\m_n^{\ 2}=\left(2M_nG+a\Sigma_n\right)^2-(1+a^2)Q_n^{\ 2}\ .
\label{mu}
\eeqn
As usual the extremal limit corresponds to sending $\m_n\rightarrow 0$, 
$\s\rightarrow \infty$ such that the charge $Q_n$ is kept fixed. Then $V=1$,
$K=0$ and $H$ is a harmonic function. 

For simplicity consider the solution representing two black holes. As before the 
rods correspond to the holes' horizons. Along the first rod the metric has the 
following limiting behavior:
\eqn
\arr{l}
\displaystyle{ds^2\sim
\cosh^{\frac{4}{1+a^2}}{\s}\left\{
f^2(\th)
\left[-\left(1-\frac{\m_1}{r}\right)\frac{dt^2}{\cosh^{\frac{8}{1+a^2}}{\s}}
+\left(\frac{\D z+\m_1+\m_2}{\D z+\m_1}\right)^2
\left(1-\frac{\m_1}{r}\right)^{-1}dr^2
\right]\right.}\\\\
\displaystyle{\ \ \ \ \ \ \ \ \ \ \ \ \ \ \ \ \ \ \ \ \ \ \ \ \ \ \left.+
\m_1^{\ 2}\left[\left(\frac{\D z+\m_1+\m_2}{\D z+\m_1}\right)^2
f^2(\th)\,d\th^2+
f^{-2}(\th)\sin^2{\th}\,d\phi^2\right]\right\}}\ ,
\earr
\label{chargedhorizon}
\eeqn
where
\eqn
f(\th)=\sqrt{\frac{\D z+\m_1(\cos{\th}+1)/2}{\D z
+\m_1(\cos{\th}+1)/2+\m_2}}\ .
\eeqn
The horizon surface gravity and area read
\eqn
\arr{c}
\displaystyle{\kappa_1=
\frac{1}{2\m_1\cosh^{\frac{4}{1+a^2}}{\s}}\,
\frac{\D z+\m_1}{\D z +\m_1+\m_2}}\ ,\\\\
\displaystyle{A(H_1)=
4\pi\m_1^{\ 2}\cosh^{\frac{4}{1+a^2}}{\s} \,
\frac{\D z +\m_1+\m_2}{\D z+\m_1}}\ .
\earr
\eeqn
Again we shall consider black holes with the same temperature $T=\kappa/(2\pi)$. 
Hence we require $\m\equiv\m_1=\m_2$ and each hole has mass $M$ and electric charge 
$Q$. The analyses of the holes' horizons is entirely similar to the Schwarzschild case.
The only difference is the conformal factor in the near-horizon metric
(\ref{chargedhorizon}).

Between the rods we have a conical singularity with deficit angle
\eqn
\frac{\d}{2\pi}=-\frac{\m^2}{(\D z+2\m)\D z}\ .
\eeqn
As before this conical singularity can be effectively described as a strut that 
connects both black holes. The corresponding energy as seen by a static observer
at infinity and the pressure on the strut along the $z$ axis are
\eqn
E=-\frac{\m^2\D z}{4G(\D z+\m)^2}\sim 
-\frac{\m^2}{4G\D z}\ ,\ \ \ 
P=\frac{\m^2}{4G(\D z+2\m)\D z}\sim 
\frac{\m^2}{4G\D z^2}\ ,
\eeqn
respectively.
The interpretation is similar to the Schwarzschild black holes case but now the 
holes interact through the gravitational, scalar and gauge fields exchange. 
This fact is revealed by formula (\ref{mu}) for the non-extremality parameter $\m$.

\subsection{Gravitational Thermodynamics}

Our starting point is the Euclidean action for the solution representing
two non-extremal charged black holes
\eqn
I=\beta
\left(2\frac{M}{2}-2\frac{Q}{8G}\Phi+\frac{\m^2\D z}{4G(\D z+\m)^2}\right)=
\beta\left(\frac{\m}{2G}+\frac{\m^2\D z}{4G(\D z+\m)^2}\right)\ ,
\eeqn
where
\eqn
\frac{1}{T}=\beta=4\pi\m\cosh^{\frac{4}{1+a^2}}{\s}\,\frac{\D z+2\m}{\D z+\m}\ ,
\ \ \ 
\Phi=\frac{2}{\sqrt{1+a^2}}\tanh{\s}\ ,
\label{temppot}
\eeqn
are the Euclidean time periodicity and the horizons electric potential, 
respectively. Then the thermodynamical potential 
$W=E-TS-\frac{Q_{total}}{4G}\Phi$ reads
\eqn
W(T,\Phi,\D z)=
\frac{\m}{2G}+\frac{\m^2\D z}{4G(\D z+\m)^2}\ ,
\eeqn
with the non-extremality parameter $\m=\m(T,\Phi,\D z)$ defined through
the relations (\ref{temppot}). The entropy $S$, generalized force $F_{\D z}$
and total charge $Q_{total}$ are determined by the usual thermodynamical
relations
\eqn
S=-\left.\frac{\partial W}{\partial T}\right|_{\Phi,\D z}\ ,\ \ \ \ 
F_{\D z}=-\left.\frac{\partial W}{\partial \D z}\right|_{T,\Phi}\ ,\ \ \ \ 
\frac{Q_{total}}{4G}=-\left.\frac{\partial W}{\partial \Phi}\right|_{T,\D z}\ .
\eeqn
After some straightforward algebra we obtained the following results:
\eqn
\arr{c}
\displaystyle{
S=\frac{8\pi\m^2}{4G}\cosh^{\frac{4}{1+a^2}}{\s}
\left[ 1+\frac{\m}{\D z}+{\cal O}\left(\frac{\m}{\D z}\right)^2\right]=
\frac{1}{4G}\left( A(H_1)+A(H_2)\right)}\ ,\\\\
\displaystyle{
F_{\D z}=-\frac{1}{4G}\left(\frac{\m}{\D z}\right)^2
\left[ 1+{\cal O}\left(\frac{\m}{\D z}\right)\right]}\ ,\\\\
\displaystyle{
Q_{total}=2Q\left[ 1+{\cal O}\left(\frac{\m}{\D z}\right)^2\right]}\ .
\earr
\eeqn
To order $\m/\D z$ the entropy follows the area law, the force $F_{\D z}$ arises
from the gravitational and scalar field attraction and 
electric repulsion between the holes and the charge $Q_{total}$
is the sum of the holes electric charge. Notice that we are using an ensemble
with the temperature and electric potential held fixed. For example, this means 
that the total mass and charge are not fixed quantities. In fact for large
separation we expect two black holes each with the same mass and charge as for 
the single black hole case ($\D z=0$).

Finally, as for the Schwarzschild black holes, 
the thermodynamical potential $W$ and the mean energy $E$ 
have an apparently wrong sign in the interacting term. However, holding the
entropy and charges fixed the variation of the energy does follow the expected
Coulomb-like behavior. We shall come back to this point in the next section.

\section{Interacting Black Holes in String/M-theory}
\news

To have a better understanding of the microscopic structure underlying the 
previous results we shall analyze the problem in String/M-theory. We consider
four-dimensional black holes with four $U(1)$ charges. A convenient 
representation of such holes is the non-extremal M-theory configuration
$5\perp 5\perp 5$ of three M5-branes intersecting over a common string along
which momentum flows \cite{KlebTsey,CvetTsey} (the M5-branes intersect pairwise
over a 3-brane). There are three magnetic charges related to the number of
M5-branes in three different hyperplanes and a electric charge which has a
Ka\l u\.{z}a-Klein origin. The four-dimensional Einstein metric for two of
such black holes aligned along the $z$ axis is
\eqn
ds^2=-H^{-\frac{1}{2}}Vdt^2+H^{\frac{1}{2}}V^{-1}\g_{ij}dx^idx^j\ ,
\eeqn
where
\eqn
H=\prod_{i=1}^4\left(1+\sinh^2{\s_i}\,(1-V)\right)\ ,
\eeqn
and the functions $V=e^{2U}$ and $K$ are given by (\ref{solution}) with $l=2$.
We shall set $\m\equiv\m_1=\m_2$ such that the black holes have the same
temperature. The M5-brane and Ka\l u\.{z}a-Klein charges associated with
each hole are 
\eqn
Q_i=\frac{\m}{2}\sinh{(2\s_i)}\ .
\eeqn
Next consider the dilute gas regime \cite{MaldStro} where the energy associated
with the Ka\l u\.{z}a-Klein excitations is much smaller than the M5-branes mass:
\eqn
\m,\ \m\sinh{\s_4}\ll 
\m\sinh{\s_j}\ ,\ \ \ \ j=1,2,3.
\eeqn
In this limit anti-branes are suppressed and the number of M5-branes $N_j$
satisfies
\eqn
\frac{e^{2\s_1}}{4}\m=
\frac{N_1}{L_6L_7}\left(\frac{\kappa_{11}}{4\pi}\right)^{2/3},\ \ \ 
\frac{e^{2\s_2}}{4}\m=
\frac{N_2}{L_4L_5}\left(\frac{\kappa_{11}}{4\pi}\right)^{2/3},\ \ \ 
\frac{e^{2\s_3}}{4}\m=
\frac{N_3}{L_2L_3}\left(\frac{\kappa_{11}}{4\pi}\right)^{2/3},
\eeqn
where the eleven-dimensional gravitational coupling satisfies
$\kappa_{11}^{\ \ 2}=V_7\kappa^2$, with $V_7$ the volume of the compact space.
The first set of M5-branes has two compact orthogonal directions with
length $L_6$ and $L_7$ and similarly for the other M5-branes. 
The mass of each black hole becomes
\eqn
M=\frac{1}{4G}\left( Q_1+Q_2+Q_3\right)+\frac{\m}{8G}\cosh{(2\s_4)}
\equiv M_{BPS}+\d M\ ,
\label{mass}
\eeqn
where $M_{BPS}$ is the total M5-branes mass and $\d M$ is the mass associated
with the Ka\l u\.{z}a-Klein excitations along the common direction with 
length $L_1$. We interpret $M_{BPS}$ as the ground state mass, while the 
Ka\l u\.{z}a-Klein modes correspond to the excitations of the system. 
The total momentum carried along this compact direction is
\eqn
P=P_L-P_R=\frac{\m}{8G}\sinh{(2\s_4)}\ .
\eeqn

In recent String Theory developments black holes are described by a dual conformal
field theory \cite{Mald,GKP,Witten}. In some cases this is a $1+1$ conformal field
theory, as in the original work \cite{StroVafa}, commonly referred as the effective
string description of black holes [15-19]. For the M-theory configuration described
above the black hole is described by an effective string with length 
$L_{eff}=N_1N_2N_3L_1$ and central charge $c=6$ \cite{KlebTsey}. Then one matches the 
energy and the momentum carried by left- and right-movers on this string with
$\d M$ and $P$, respectively. As a result the entropy formula is recovered 
exactly. In the case at hand each black hole will still be described by
an effective string. However, the interaction between both black holes means that
the strings will interact with each other and that the level for left- and 
right-movers will change. To make these ideas precise we write the thermodynamical 
potential $W$ and the entropy of the system in the dilute gas regime and for large 
distance between the holes
\eqn
\arr{c}
\displaystyle{W=\frac{\m}{2G}+\frac{\m^2}{4G\D z}}\ ,\\\\
\displaystyle{S=\frac{8\pi}{4G}\sqrt{Q_1Q_2Q_3}\sqrt{\m}\cosh{\s_4}
\left(1+\frac{\m}{\D z}\right)}\ .
\earr
\label{WS}
\eeqn
Then the energy mean value $E=W+TS+\frac{Q_{total}}{4G}\Phi$ reads
\eqn
E=2\left( M_{BPS}+\d M+\frac{\m^2}{4G\D z}\right)-\frac{\m^2}{4G\D z}
\equiv 2\left( M_{BPS}+\d E\right)+V_{int}\ ,
\eeqn
where $M_{BPS}$ and $\d M$ are defined in (\ref{mass}). We see that in this
case the increase in energy $\d E$ does not coincide with the increase in the
mass $\d M$ of each hole. As we start filling the left- and right-moving states
in each effective string, they will interact reducing their energy by $V_{int}$
according to a large distance attractive Coulomb potential. In fact, the dependence 
on the string coupling $V_{int}\sim -g^2N_L N_R$ suggests that in a similar D-brane
setting (e.g. the $D5-D1$ system for five-dimensional black holes) the interactions
arise from an annulus diagram with two insertions on a boundary. 
Thus, in order to reproduce a state corresponding to two black holes each of mass 
$M_{BPS}+\d M$ we need an energy 
associated with the string excitations given by $\d E=\d M+\frac{\m^2}{4G\D z}$.
This fact explains the apparent discrepancy in the sign of the interacting
term in the thermodynamical potential $W$ and energy $E$. There are two different
contributions: one from the Coulomb attraction and another from the black holes
internal states. In Figure 4 we represent the effect of the interaction 
between the effective strings. 

\begin{figure}
\centerline{\psfig{figure=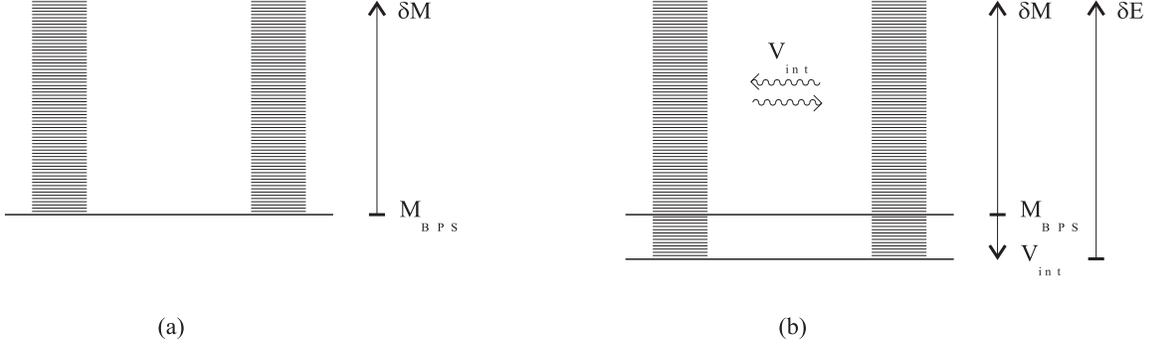,width=6in}}
\caption{\small{Representation of the tower of momentum states on the effective 
string model for both black holes when their mutual interaction is (a) switched
off and (b) switched on. The effect of the Coulomb attraction on the effective string 
states is to shift their ground state energy by $V_{int}$.}}
\label{fig4}
\end{figure}

To determine the level of the left- and right-moving states 
on each effective string note that
\eqn
\arr{c}
\displaystyle{
\d E=\frac{1}{4G}\left[\frac{\m\cosh{2\s_4}}{2}+\frac{\m^2}{\D z}\right]
=\frac{1}{4G}\left[\left(\frac{e^{2\s_4}}{4}+\frac{\m}{2\D z}\right)\m
+\left(\frac{e^{-2\s_4}}{4}+\frac{\m}{2\D z}\right)\m\right]}\ ,\\\\
\displaystyle{
P=\frac{1}{4G}\frac{\m\sinh{2\s_4}}{2}
=\frac{1}{4G}\left[\left(\frac{e^{2\s_4}}{4}+\frac{\m}{2\D z}\right)\m
-\left(\frac{e^{-2\s_4}}{4}+\frac{\m}{2\D z}\right)\m\right]}\ .
\earr
\eeqn
Hence there is a shift in the left- and right-mover levels. Now we have
\eqn
\kappa^2\frac{N_L}{L_1}=
\left(\frac{e^{2\s_4}}{4}+\frac{\m}{2\D z}\right)\m\ ,\ \ \ 
\kappa^2\frac{N_R}{L_1}=
\left(\frac{e^{-2\s_4}}{4}+\frac{\m}{2\D z}\right)\m\ .
\eeqn
Then the entropy associated with both effective strings is 
\eqn
S=2\cdot 2\pi\sqrt{N_1N_2N_3}\left(\sqrt{N_L}+\sqrt{N_R}\right)\ ,
\eeqn
which matches exactly the result in (\ref{WS}), including the correction of 
order $\m/\D z$. Thus we 
successfully reproduced the deviations in the entropy formula due to the
long range interactions between these non-extremal charged black holes. 

\section{Conclusion}

In this paper we analysed in detail the geometry representing $l$ collinear 
non-extremal black holes. In spite of the conical singularity connecting the black
holes, this solution has a well defined action. This fact enabled us to study the
interaction between largely separated non-extremal black holes by using standard 
gravitational thermodynamics techniques. Then we analysed this problem using the
modern String Theory language and successfully reproduced the entropy formula 
for interacting black holes, including the correction due to the long range
force between the holes. This interaction causes the left- and right-moving 
states that build up the internal degrees of freedom of the system to reduce 
their energy affecting the state counting. 

It is interesting to note that the the binding energy due to the interaction
between the holes is expected to lead to a energy shift in the observed Hawking
radiation. In fact left- and right-movers with quantum numbers $n_{L,R}$ such
that $\frac{n_{L,R}}{R_1}<-V_{int}$ are bounded and cannot leave the branes. 
Unfortunately, it may be very difficult to see this by solving Laplace equation (e.g.
for a scalar field) in this background. For example, the same problem in the
simpler double-centred extremal case turns out to be highly non-trivial 
\cite{RashVisw,Costa}.

A point that is less clear to us is related to the quadratic and higher order 
corrections in $\m/\D z$ to the thermodynamical quantities here calculated. In 
particular the entropy does not follow the area law to this order. We suspect
that the conical singularity is a good approximation to the long range Coulomb
interaction but breaks down otherwise. Indeed, we saw that for small distances
between the black holes the physics is Planckian and our approximations are
no longer valid. 

A possible venue of research is the study of this geometry in the decoupling
limit. This may lead to interesting results on the Coulomb branch of the
dual field theory at finite temperature (which is not stable since the 
Coulomb branch is lifted). For example, a study of the $AdS/CFT$ duality for
${\cal N}=4$ Super-Yang-Mills at finite temperature using D3-brane probes can
be found in \cite{Kiri}. Of course one may wish first to generalize the above 
collinear non-extremal charged black hole solutions to arbitrary space-time 
dimensions.

\section*{Acknowledgments}

We would like to thank Frank Ferrari and Lori Paniak for many discussions.  
MSC is supported by FCT (Portugal) under programme PRAXIS XXI and by 
the NSF grant PHY-9802484.

\end{document}